\documentclass[useAMS,usenatbib,usegraphicx]{mn2e}

\usepackage{amssymb}

\newcommand\aj{{AJ}}
\newcommand\apj{{ApJ}}
\newcommand\apjl{{ApJ}}
\newcommand\apss{{Ap\&SS}}
\newcommand\aap{{A\&A}}
\newcommand\mnras{{MNRAS}}
\newcommand\pasp{{PASP}}
\newcommand\pasj{{PASJ}}

\newcommand{\actaa}{Acta Astron.}      

\title[The AM CVn binary SDSS\,J1730+5545]{The AM CVn binary SDSS\,J173047.59+554518.5}
\author[P. J. Carter et al.]{P. J. Carter,$^{1}$\thanks{E-mail: philip.carter@warwick.ac.uk} D. Steeghs,$^{1}$ T. R. Marsh,$^{1}$ T. Kupfer,$^{2}$ C. M. Copperwheat,$^{3}$ \newauthor P. J. Groot$^{2}$ and G. Nelemans$^{2,4}$\\
$^{1}$Department of Physics, University of Warwick, Coventry CV4 7AL\\
$^{2}$Department of Astrophysics/IMAPP, Radboud University Nijmegen, PO Box 9010, 6500 GL Nijmegen, the Netherlands\\
$^{3}$Astrophysics Research Institute, Liverpool John Moores University, IC2, Liverpool Science Park, 146 Brownlow Hill, Liverpool L3 5RF\\
$^{4}$Institute for Astronomy, KU Leuven, Celestijnenlaan 200D, 3001 Leuven, Belgium\\
}

\begin{document}

\date{Accepted 2013 October 28.}

\pagerange{\pageref{firstpage}--\pageref{lastpage}} \pubyear{2013}

\maketitle

\label{firstpage}

\begin{abstract}
The AM Canum Venaticorum (AM CVn) binaries are a rare group of hydrogen-deficient, ultra-short period, mass-transferring white dwarf binaries, and are possible progenitors of type Ia supernovae.
We present time-resolved spectroscopy of the recently-discovered AM CVn binary SDSS\,J173047.59+554518.5. The average spectrum shows strong double-peaked helium emission lines, as well as a variety of metal lines, including neon; this is the second detection of neon in an AM CVn binary, after the much brighter system GP Com. We detect no calcium in the accretion disc, a puzzling feature that has been noted in many of the longer-period AM CVn binaries. We measure an orbital period, from the radial velocities of the emission lines, of 35.2\,$\pm$\,0.2\,minutes, confirming the ultra-compact binary nature of the system. The emission lines seen in SDSS\,J1730 are very narrow, although double-peaked, implying a low inclination, face-on accretion disc; using the measured velocities of the line peaks, we estimate $i\,\le11 ^{\circ}$. This low inclination makes SDSS\,J1730 an excellent system for the identification of emission lines.\end{abstract}



\section{Introduction}

The AM Canum Venaticorum (AM CVn) binaries are a class of ultracompact systems, characterised by their short orbital periods, which range from 5 to 65\,minutes, and their hydrogen-deficient spectra. They consist of a white dwarf accreting helium-rich material from a (semi-)degenerate companion. It is the degenerate nature of the mass donor that allows their periods to lie well below the believed minimum period ($\sim$80\,min) of hydrogen-rich cataclysmic variables (CVs; \citealt{1982ApJ...254..616R,2009MNRAS.397.2170G}). A recent review of the AM CVn binaries is given by \citet{2010PASP..122.1133S}.

The evolution of AM CVn binaries is thought to be driven by gravitational wave radiation. Combined with the increasing period and decreasing mass ratio, this gives rise to a very strong dependence of the accretion rate, and hence their observed properties, on the orbital period.
In the shortest period systems ($P_{\rmn{orb}} \lesssim$ 10\,min) the accretion stream likely impacts directly onto the surface of the accretor, and no accretion disc forms \citep{2002MNRAS.331L...7M,2010ApJ...711L.138R}. Systems with orbital periods of $\sim$10 to $\sim$20\,minutes possess an accretion disc that is in a stable `high' state, and their spectra are dominated by helium absorption from the optically thick disc \citep{1994MNRAS.271..910O}. For the longest period systems ($P_{\rmn{orb}} \gtrsim$ 45\,min) the disc is in a stable low state; photometry of these systems typically lacks orbital signatures, and their optical spectra are dominated by helium emission lines \citep{1995Ap&SS.225..249W,2001ApJ...552..679R}. In the intermediate period systems (20 $\lesssim P_{\rm orb}\lesssim$ 45\,min) the disc is unstable, and their appearance varies between that of the high-state and the low-state systems, analogously to the hydrogen-rich dwarf novae \citep{1997PASJ...49...75T,2012A&A...544A..13K,2012MNRAS.419.2836R}.

The AM CVn binaries are an extremely important class of objects due to the strong low-frequency gravitational wave emission that governs their evolution \citep{2006CQGra..23S.809S,2006MNRAS.371.1231R,2009CQGra..26i4030N,2012ApJ...758..131N}. AM CVn binaries are a possible progenitor for Type Ia supernovae via the `double detonation' mechanism \citep[e.g.][]{2009ApJ...699.1365S,2013arXiv1305.6925S}, and may produce rare sub-luminous SN Ia-like events (`SN.Ia'; \citealt{2007ApJ...662L..95B,2011MNRAS.411L..31B}).

The first AM CVn binaries were discovered serendipitously in a variety of different surveys. This made it impossible to study their population and to derive fundamental quantities required for calibration of predictions from binary evolution theory.
In the past ten years dedicated surveys have tripled the number of known AM CVn binaries [samples based on outburst selection target systems with orbital periods in the 20 -- 45\,min range \citep{2011ApJ...739...68L,2013MNRAS.430..996L}, whilst the samples based on the Sloan Digital Sky Survey \citep[SDSS;][]{2000AJ....120.1579Y} select low-state systems \citep{2005MNRAS.361..487R,2005AJ....130.2230A,2008AJ....135.2108A,2009MNRAS.394..367R,2010ApJ...708..456R,2013MNRAS.429.2143C}]. These homogeneous samples have allowed the first detailed studies of the population, and estimation of the AM CVn space density \citep{2007MNRAS.382..685R,2013MNRAS.429.2143C}.

Characterisation of the sample requires detailed follow-up observations after the initial discovery, in order to determine the orbital period and confirm the ultracompact binary nature of these candidate systems. Here we present time-resolved spectroscopy of the candidate AM CVn binary SDSS\,J173047.59+554518.5 (hereafter SDSS\,J1730), identified spectroscopically by \citet{2013MNRAS.429.2143C} amongst a colour-selected sample of objects from SDSS photometry.


\section{Observations and data reduction}

We obtained optical spectroscopy of SDSS\,J1730 on 2012 May 21 with the Gemini Multi-Object Spectrograph (GMOS; \citealt{2004PASP..116..425H}) at the Gemini-North telescope on Mauna Kea, Hawaii. The observations consist of 54 spectra, obtained using the B600+ grating with a 180\,s exposure time. GMOS has three 2048$\times$4608 e2v deep depletion CCDs, which were used in six amplifier mode. The resulting spectra cover the wavelength range 4120\,--\,6973\,\AA\, with a resolution of $\sim$5.46\,\AA.

The GMOS observations used a 1.0 arcsec slit; 4$\times$4 binning and `slow' readout mode were used to minimise the read-out noise.

\begin{table}
\centering
\caption{Log of our observations of SDSS\,J1730.}
\label{t:speclog}
\begin{tabular}{l l r r}
\hline
Date		& UT		& Exposure	& Exposures \\
		&		& time (s)	&	\\
\hline
Gemini/GMOS	& & & \\
2012 May 21	& 10:45--13:34	& 180		& 54 \\
 & & & \\
WHT/ISIS	& & & \\
2012 July 13	& 22:21--03:16	& 300		& 54 \\
2012 July 14	& 21:14--03:24	& 300		& 67 \\
2012 July 15	& 23:09--03:28	& 300		& 42 \\
\hline
\end{tabular}
\end{table}

We obtained a second, longer set of phase-resolved spectra of SDSS\,J1730 on 2012 July 13, 14 and 15 on the William Herschel Telescope (WHT), situated on the island of La Palma, with the Intermediate dispersion Spectrograph and Imaging System (ISIS). These observations consist of 163 spectra obtained with 300\,s exposures using the R600B grating in the blue arm and the R600R grating in the red arm. The resulting spectra cover the wavelength range 3750--5260\,\AA\ and 5540--7125\,\AA, with a spectral resolution of $\sim$2.43\,\AA\ in the blue arm, and $\sim$2.18\,\AA\ in the red arm.

All ISIS observations used 4$\times$2 binning and a 1.2 arcsec slit. The detectors were used in `slow' readout mode to minimise the read-out noise. The slit was kept at a fixed position angle of 145.3$^{\circ}$ in order to observe both SDSS\,J1730 and a nearby comparison star (SDSS\,J173050.53+554442.5, 44\,arcsec south east of SDSS\,J1730) simultaneously.

Bias subtraction was achieved using average bias frames constructed from bias frames taken each night. Normalized flatfield frames were constructed by averaging a series of tungsten lamp flatfield frames obtained during each night, additionally, for the ISIS observations, spatial profile correction was achieved using nightly averaged twilight flatfields.

Wavelength calibration was obtained from CuAr arc lamp exposures taken at the start, middle and end of the three hour GMOS observing block, and CuNeAr arc exposures obtained every $\sim$40\,minutes for the duration of our ISIS observations of SDSS\,J1730.
Fourth order polynomial fits to reference lines resulted in $\sim$0.15\,\AA\ and $\sim$0.05\,\AA\ rms residuals for the GMOS and ISIS spectra. Each exposure was wavelength calibrated by interpolating the solutions obtained from the two closest arcs, in order to correct for instrument flexure.

The comparison star spectra were used to correct the ISIS spectra of SDSS\,J1730 for time and wavelength dependent slit-losses. The individual comparison star spectra were divided by the master comparison star spectrum obtained at lowest air mass. These ratio spectra were fitted with a second order polynomial, and the spectra of SDSS\,J1730 corrected by dividing by the corresponding polynomial fit. The ISIS spectra were flux-calibrated and corrected for instrumental response using the spectrophotometric standard star BD+28\,4211.

All spectra were reduced using optimal extraction as implemented in the \textsc{pamela}\footnote{\textsc{pamela} is included in the \textsc{starlink} distribution `Hawaiki' and later releases. The \textsc{starlink} Software Group homepage can be found at http://starlink.jach.hawaii.edu/starlink.} code \citep{1989PASP..101.1032M}, which also uses the \textsc{starlink} packages \textsc{kappa}, \textsc{figaro} and \textsc{convert}. Spectra were wavelength and flux calibrated using \textsc{molly}\footnote{\textsc{molly} was written by T. R. Marsh and is available from http://www.warwick.ac.uk/go/trmarsh/software.}. Our spectroscopic observations are summarised in Table \ref{t:speclog}.


\section{Results}

\subsection{Average spectral features}

%
\begin{figure*}
 \includegraphics[width=1.0\textwidth]{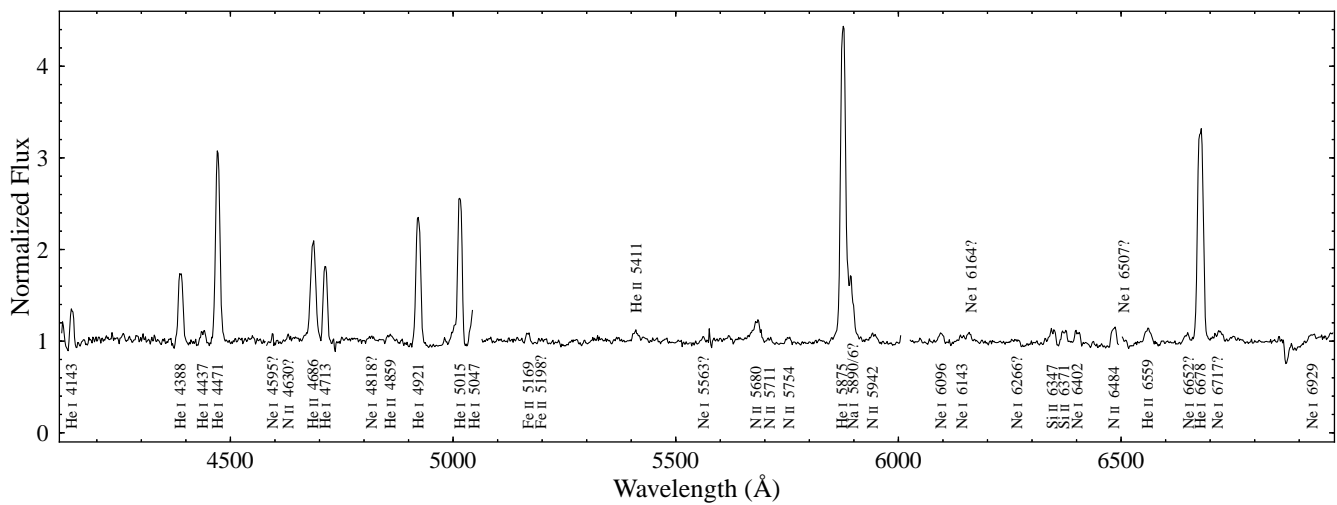}
 \caption{Normalized average spectrum of SDSS\,J1730 obtained with GMOS. The strong helium lines are prominent, in addition many weak metal lines are detected. The small gaps at 5055 and 6015\,\AA\ are due to gaps between the individual CCDs.\label{f:gemspec}}
\end{figure*}
\begin{figure*}
 \includegraphics[width=1.0\textwidth]{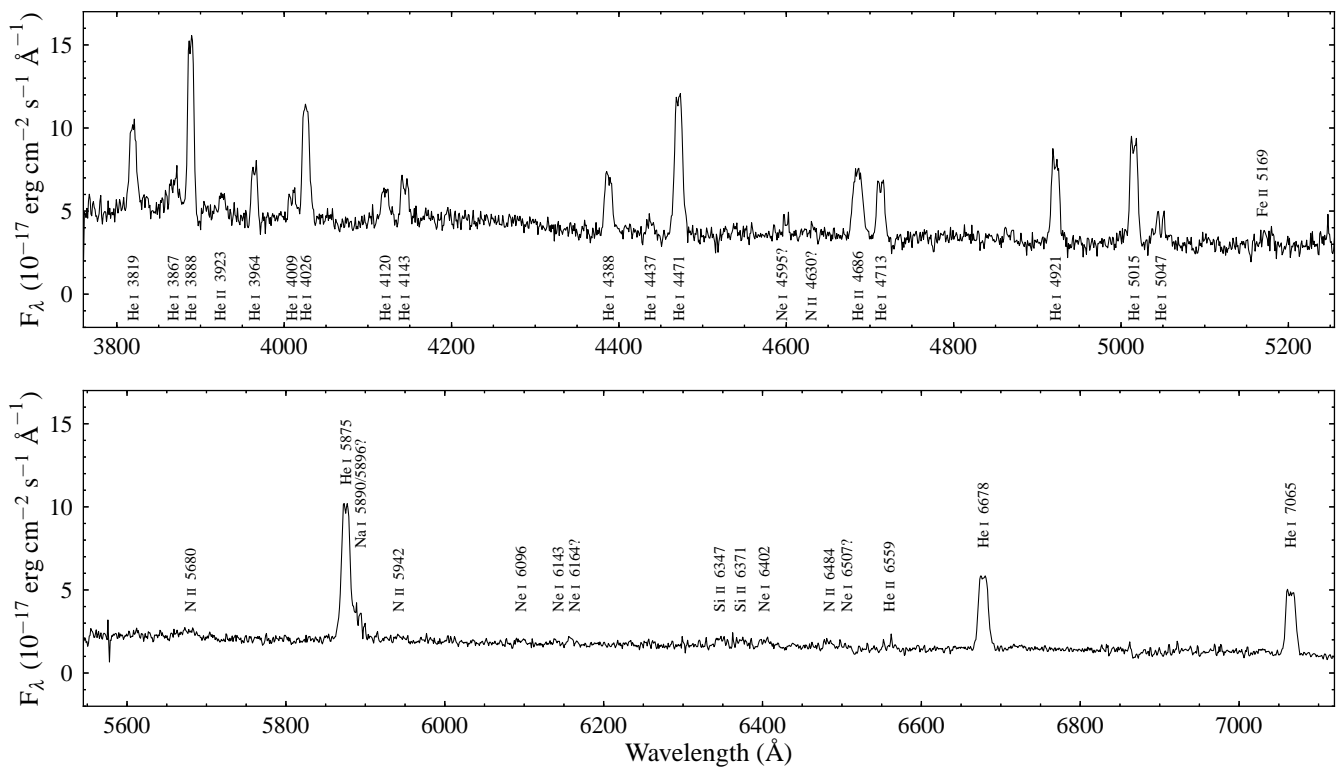}
 \caption{Average spectrum of SDSS\,J1730 obtained with ISIS. The prominent lines have been labelled. \label{f:avspec}}
\end{figure*}
The average GMOS spectrum of SDSS\,J1730 is shown in Fig. \ref{f:gemspec}. It shows strong, narrow He\,\textsc{i} emission lines, as well as several He\,\textsc{ii} lines. Given the strength of the He\,\textsc{ii} 5411\,\AA\ line, and the absence of H$\gamma$, we associate the emission close to 6559 and 4859\,\AA\ with He\,\textsc{ii}, and not with the hydrogen Balmer series. There is no clear sign of the accreting white dwarf in the spectrum.

Several lines of Fe, N, Si and Ne are also identified. Metal lines are often seen in quiescent AM CVn spectra \citep[e.g.][]{2006MNRAS.365.1109R,2009MNRAS.394..367R,2013MNRAS.430..996L,2013MNRAS.432.2048K}, however, neon has only previously been identified in GP Com \citep{1991ApJ...366..535M,2010MNRAS.401.1347N}.

The ISIS spectrum (Fig. \ref{f:avspec}) shows a blue continuum, and reveals the double-peaked line profiles expected from the accretion disc. The red wing of the He\,\textsc{i} 5875 line shows contamination from an emission feature around 5893\,\AA\ which we identify as being caused by the Na D doublet. \citet{2006MNRAS.365.1109R} reported similar blending in the spectra of SDSS\,J1240-0159 and V406 Hya. This feature is puzzling as any sodium in the disc should be largely ionized at the temperatures required to produce most of the emission lines observed \citep[see][for further discussion]{1991ApJ...366..535M}.

Table \ref{t:ew} shows the equivalent width (EW) and FWHM of the prominent lines in the average spectra of SDSS\,J1730. The narrow emission lines, that reveal their double-peaked nature clearly in the higher resolution ISIS spectra, imply a low inclination. This also results in a low radial velocity amplitude, which makes the periodic motion of the binary difficult to measure. \citet{2013MNRAS.429.2143C} used the EW of the He\,\textsc{i} 5875 line to predict an orbital period of 40--55\,minutes based on the known population of AM CVn binaries.

\begin{table*}
\centering
\begin{minipage}{110mm}
\caption{Equivalent widths and FWHM of prominent lines.}
\label{t:ew}
\begin{tabular}{l r r r r}
\hline
Line			& \multicolumn{2}{c}{GMOS} 		& \multicolumn{2}{c}{ISIS} \footnotetext{\hspace{-2mm}Note: Values marked `\ldots' could not be measured reliably. Values marked `--' indicate that the line falls outside the wavelength coverage.} \\
			& EW (\AA{})		& FWHM (km s$^{-1}$)	& EW (\AA{})		& FWHM (km s$^{-1}$) \\
\hline
He\,\textsc{i} 3819	& --			& --			& $-$9.2\,$\pm$\,0.7	& 670\,$\pm$\,40 \\
He\,\textsc{i} 3867	& --			& --			& $-$4.6\,$\pm$\,0.7	& 970\,$\pm$\,130 \\
He\,\textsc{i} 3888	& --			& --			& $-$16.3\,$\pm$\,1.0	& 550\,$\pm$\,30 \\
He\,\textsc{i} 3964	& --			& --			& $-$4.4\,$\pm$\,0.7	& 480\,$\pm$\,60 \\
He\,\textsc{i} 4026	& --			& --			& $-$14.3\,$\pm$\,0.7	& 650\,$\pm$\,30 \\
He\,\textsc{i} 4120	& --			& --			& $-$3.4\,$\pm$\,1.0	& 630\,$\pm$\,150 \\
He\,\textsc{i} 4143	& $-$3.1\,$\pm$\,0.4	& 550\,$\pm$\,50	& $-$5.0\,$\pm$\,0.8	& 670\,$\pm$\,80 \\
He\,\textsc{i} 4388	& $-$8.3\,$\pm$\,0.4	& 680\,$\pm$\,20	& $-$9.2\,$\pm$\,0.7	& 620\,$\pm$\,30 \\
He\,\textsc{i} 4437	& $-$1.2\,$\pm$\,0.4	& 670\,$\pm$\,170	& $-$1.7\,$\pm$\,0.6	& 600\,$\pm$\,170 \\
He\,\textsc{i} 4471	& $-$23.7\,$\pm$\,0.4	& 705\,$\pm$\,10	& $-$24.4\,$\pm$\,0.7	& 660\,$\pm$\,20 \\
He\,\textsc{ii} 4686	& $-$15.1\,$\pm$\,1.2	& 830\,$\pm$\,50 	& $-$13.7\,$\pm$\,1.3	& 730\,$\pm$\,50 \\
He\,\textsc{i} 4713	& $-$9.0\,$\pm$\,0.3	& 620\,$\pm$\,20	& $-$10.3\,$\pm$\,0.7	& 610\,$\pm$\,30 \\
He\,\textsc{i} 4921	& $-$16.6\,$\pm$\,0.4	& 650\,$\pm$\,10	& $-$17.6\,$\pm$\,0.8	& 620\,$\pm$\,20  \\
He\,\textsc{i} 5015	& $-$18.4\,$\pm$\,0.4	& 610\,$\pm$\,10 	& $-$22.6\,$\pm$\,0.9	& 590\,$\pm$\,20  \\
He\,\textsc{i} 5047	& \ldots		& \ldots		& $-$8.0\,$\pm$\,1.0	& 920\,$\pm$\,90 \\
Fe\,\textsc{ii} 5169	& $-$0.9\,$\pm$\,0.2	& 470\,$\pm$\,70	& $-$3.0\,$\pm$\,1.0	& 750\,$\pm$\,190 \\
He\,\textsc{ii} 5411	& $-$1.7\,$\pm$\,0.3	& 860\,$\pm$\,100	& --			& --	 \\
N\,\textsc{ii} 5680	& $-$4.4\,$\pm$\,0.4	& 1050\,$\pm$\,70	& $-$4.4\,$\pm$\,1.1	& 1200\,$\pm$\,200 \\
He\,\textsc{i} 5875	& $-$47.1\,$\pm$\,1.1	& 660\,$\pm$\,10	& $-$55.8\,$\pm$\,1.3	& 640\,$\pm$\,10 \\
Ne\,\textsc{i} 6096	& $-$1.1\,$\pm$\,0.3	& 510\,$\pm$\,90	& $-$1.1\,$\pm$\,0.6	& 500\,$\pm$\,200 \\
Si\,\textsc{ii} 6347	& $-$2.3\,$\pm$\,0.3	& 670\,$\pm$\,60	& $-$3.3\,$\pm$\,0.7	& 570\,$\pm$\,100 \\
Si\,\textsc{ii} 6371	& $-$2.3\,$\pm$\,0.3	& 660\,$\pm$\,60	& $-$3.7\,$\pm$\,1.0	& 820\,$\pm$\,160 \\
Ne\,\textsc{i} 6402	& $-$2.0\,$\pm$\,0.3	& 610\,$\pm$\,60	& $-$3.0\,$\pm$\,0.9	& 630\,$\pm$\,150 \\
N\,\textsc{ii} 6484	& $-$2.3\,$\pm$\,0.3	& 490\,$\pm$\,50	& $-$2.8\,$\pm$\,0.7	& 470\,$\pm$\,90 \\
He\,\textsc{ii} 6559	& $-$2.8\,$\pm$\,0.2	& 730\,$\pm$\,50	& $-$3.1\,$\pm$\,1.0	& 610\,$\pm$\,160 \\
He\,\textsc{i} 6678	& $-$34.8\,$\pm$\,0.6	& 590\,$\pm$\,10	& $-$44.1\,$\pm$\,1.1	& 560\,$\pm$\,10 \\
He\,\textsc{i} 7065	& --			& --			& $-$48.9\,$\pm$\,1.8	& 540\,$\pm$\,10 \\
\hline
\vspace{-1cm}
\end{tabular}
\end{minipage}
\end{table*}

We fit the average flux-calibrated spectrum with a blackbody, masking the obvious emission lines, giving a continuum temperature $T$\,=\,12\,700\,$\pm$\,100\,K. We give here the formal uncertainty in the fit, note that this does not reflect the considerable uncertainties in the contributions of the white dwarf and the disc to the continuum.
The data were corrected for Galactic extinction using the $R_V=3.1$ prescription of \citet{1999PASP..111...63F} and the full Galactic reddening according to \citet{1998ApJ...500..525S}, $E(B-V) = 0.047$. Without correction for extinction we obtain a temperature of 11\,600\,K. 
\begin{figure}
 \includegraphics[width=0.48\textwidth]{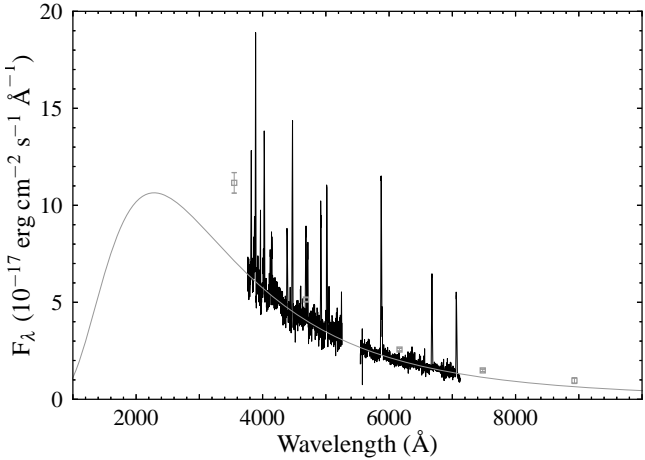}
 \caption{Flux calibrated ISIS average spectrum of SDSS\,J1730 together with SDSS photometric fluxes (grey squares). The grey line overplotted is a blackbody fit to the spectrum, $T$\,=\,12\,700\,$\pm$\,100\,K. Data have been corrected for Galactic extinction.\label{f:BBfit}}
\end{figure}
%

\subsection{The spectroscopic period}

To find the spectroscopic period we measured the radial velocity variation of the He\,\textsc{i} emission lines  using \textsc{molly}. These lines, formed in the disc, should approximately trace the motion of the accretor via the orbital motion of the accretion disc. For measurement of the orbital period, all we require is some signal in the radial velocities of the lines that varies on the orbital period.

Each line was fit by a single Gaussian, with the initial values of the fit parameters determined from a fit to the average spectrum. The FWHM for each line was fixed to this value, whilst the velocity offset common to all lines, and the height of each line were allowed to vary. We use the 6 strongest lines excluding He\,\textsc{i} 5875 (which causes a significant offset, probably due to contamination from the nearby Na feature) for the GMOS spectra, and the 11 strongest He\,\textsc{i} lines for the ISIS spectra. The fitting procedure thus yields, for each spectrum, a single value for the radial velocity, with a much greater precision than could be obtained for any single line. This allows us to detect the low amplitude velocity variations exhibited by SDSS\,J1730.

We constructed a radial velocity curve from these measurements, and calculated the Lomb-Scargle periodogram \citep{1982ApJ...263..835S}; we assume that the strongest peak corresponds to the orbital period. The radial velocities were then fit with a circular orbit of the form,
\begin{equation}\label{e:rv}
V(t) \, = \, K \; \rmn{sin}\left(2\pi(t-\rmn{HJD}_0) \over P_{\rmn{orb}} \right) \; + \; \gamma,
\end{equation}
where $P_{\rmn{orb}}$ is fixed to the value derived from the strongest peak in the periodogram.

\begin{figure}
 \includegraphics[width=0.48\textwidth]{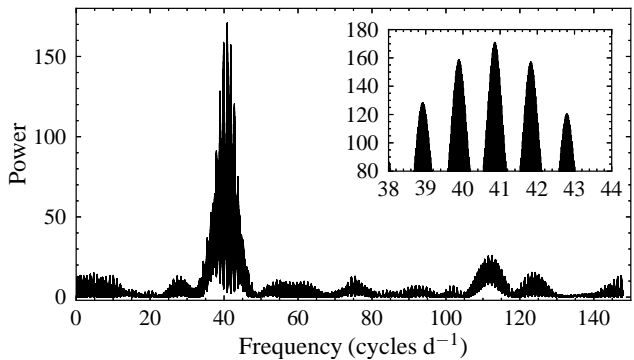}
 \caption{Lomb-Scargle periodogram calculated from the He\,\textsc{i} radial velocities of SDSS\,J1730. \label{f:pgram}}
\end{figure}

We show the Lomb-Scargle periodogram calculated from the He\,\textsc{i} radial velocities in Fig. \ref{f:pgram}. A clear signal is seen around 40.86\,cycle\,d$^{-1}$, corresponding to a period of 35.2\,minutes. As well as the obvious $\pm$1\,cycle\,d$^{-1}$ aliases, there is a fine alias structure caused by the $\sim$2\,month offset between the Gemini and WHT observations.
\begin{figure}
 \includegraphics[width=0.48\textwidth]{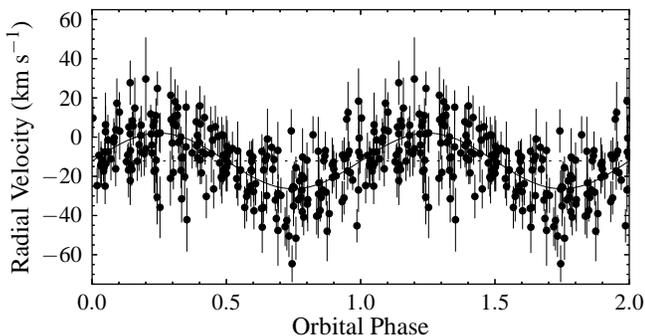}
 \caption{Measured He\,\textsc{i} radial velocities folded on a period of 35.2\,minutes. The solid and dotted lines are the best fit radial velocity curve and $\gamma$ velocity, the parameters are shown in Table \ref{t:RVfits}. \label{f:4686rv}}
\end{figure}
The He\,\textsc{i} radial velocity curve, folded on this period, is shown in Fig. \ref{f:4686rv}. The parameters of the fits to Equation \ref{e:rv}, derived from the radial velocity measurements, are shown in Table \ref{t:RVfits}. We take the zero phase, HJD$_0$, as being defined by the blue to red crossing of the velocities.

We tried various sets of lines, finding that the exact choice has no effect on the period, we therefore use the maximum possible number of (strong) lines. The Lomb-Scargle derived period was also compared to the value derived using the phase dispersion minimisation method \citep{1978ApJ...224..953S}, which gave consistent results. We cannot rule out some distortion of the systemic velocity due to the contamination of the He\,\textsc{i} 5875 line in the ISIS spectra, however, we find consistent values for $\gamma$ when including or excluding this line, and so we include it for the ISIS radial velocity measurements in order to increase the S/N.
\begin{table*}
\begin{minipage}{80mm}
\centering
\caption{Orbit parameters derived from radial velocity measurements. The orbital period is taken as the strongest peak in the periodogram, and its uncertainty derived using the bootstrap method. The zero phase, velocity amplitude and systemic velocity, and their corresponding uncertainties, result from the fit to equation \ref{e:rv}.}
\label{t:RVfits}
\begin{tabular}{c c c c}
\hline
$P_{\rmn{orb}}$ (min)		& HJD$_0$		& $K$ (km\,s$^{-1}$)	& $\gamma$ (km\,s$^{-1}$) \\
\hline
35.2\,$\pm$\,0.2		& 2456068.9549(3)	& 14.2\,$\pm$\,1.1		& $-$12.2\,$\pm$\,0.8 \\
\hline\end{tabular}
\end{minipage}
\end{table*}

The uncertainty on the period was estimated by carrying out 1000 bootstrap selections \citep{1979AnnStatist.bootstrap} of the radial velocity curve. For each subset, 217 radial velocities were selected from the full radial velocity curve, allowing for points to be selected more than once, and the periodogram calculated, taking the strongest peak as the period. The standard deviation of these computed periods, 0.2\,min, is taken as a measure of the uncertainty in the derived orbital period.

We note that there is significant power in the $\pm$1\,cycle\,d$^{-1}$ aliases seen in Fig. \ref{f:pgram}, which correspond to periods of 34.4 and 36.1\,minutes (see also next section).

\subsection{Dynamic spectrum}

%
\begin{figure*}
 \includegraphics[width=1.0\textwidth]{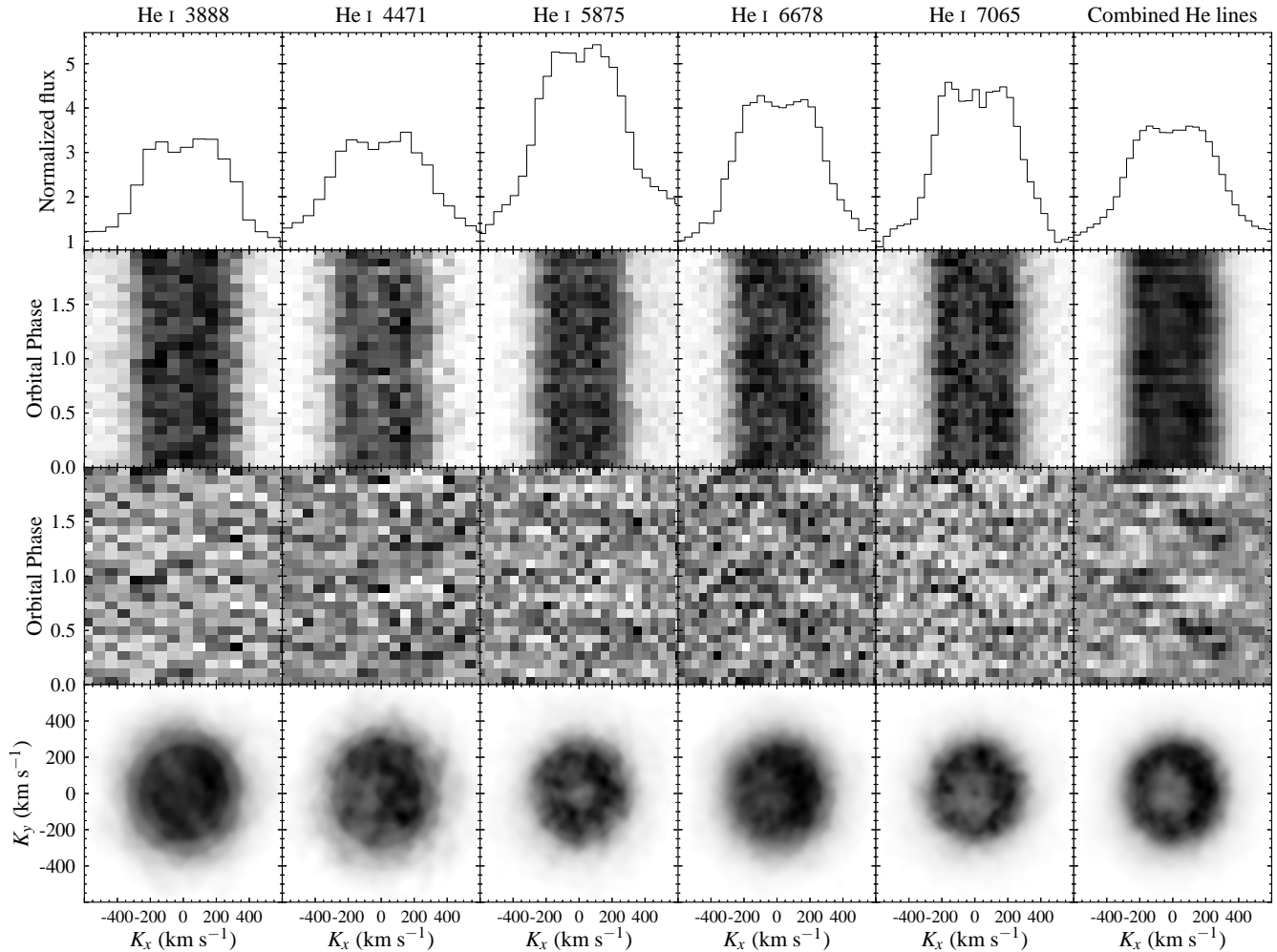}
 \caption{Average line profiles, phase-folded trailed spectra and Doppler maps for the strongest lines, and a combination of the 8 strongest He\,\textsc{i} lines. Here we use only the higher resolution ISIS spectra. The second row of trailed spectra have the average line profile subtracted in order to enhance the variability.\label{f:trail}}
\end{figure*}
Trailed spectra for the strongest lines, folded on the orbital period, 35.2\,minutes (40.9\,cycles\,d$^{-1}$), are shown in Fig. \ref{f:trail}. 
Some variability is evident in the trailed spectra, but the S-wave -- commonly caused by the bright spot, formed by the impact of the accretion stream onto the outer disc -- is clearly discernible only in the He\,\textsc{i} 6678, 7065 and combined He\,\textsc{i} line trails. The trailed spectra are also displayed with the average line profile subtracted in order to enhance the variability. The corresponding Doppler tomograms \citep{1988MNRAS.235..269M} give a similar picture of emission in the disc. A bright region or spot is visible in most of the lines shown. Since these maps were calculated using the assumed zero phase, HJD$_0$\,=\,2456068.9549, the maps may be rotated due to the unknown phase shift between our zero phase and the true zero phase of the white dwarf.

Doppler maps created at the frequency of the strongest peak in the periodogram and on the daily aliases show a weak preference for the central frequency, however, we are unable to rule out the $\pm$1\,cycle\,d$^{-1}$ aliases due to the lack of a strong bright spot.


\section{Discussion}

We measure an orbital period $P_{\rmn{orb}}\,=\,35.2\,\pm\,0.2$\,minutes, confirming SDSS\,J1730 as an ultracompact binary. This puts SDSS\,J1730 in the longer period region of the outbursting AM CVn binary period distribution, and we would expect that the system would spend the majority of its time in a low state (with appearance similar to that observed here), with infrequent, longer-duration outbursts, similar to those observed in V406 Hya ($P_{\rmn{orb}}\,=\,33.8$\,min; \citealt{2006MNRAS.365.1109R}) and SDSS\,J0129 ($P_{\rmn{orb}}\,=\,37.6$\,min; \citealt{2013MNRAS.432.2048K}) \citep{2012MNRAS.419.2836R}. We note that the Catalina Real-Time Transient Survey lightcurve appears to show the decline from a single outburst during its 7 year coverage \citep{2009ApJ...696..870D}. The measured period is slightly shorter than suggested by the EW of the He\,\textsc{i} 5875 emission line \citep{2013MNRAS.429.2143C}, but does not indicate a significant deviation from the trend of EW increasing with orbital period.

Using double Gaussian fits to the emission lines we find an average peak separation of 320\,km\,s$^{-1}$. The velocities of the two peaks in the line profile produced by an accretion disc are related to the Keplerian velocity at the outer radius of the disc, and the inclination \citep{1981AcA....31..395S,1986MNRAS.218..761H}. Thus the velocities estimated for the peaks, $\pm$\,160\,km\,s$^{-1}$, can be used to constrain the inclination of the disc in SDSS\,J1730, with some assumptions about the masses of the two components, and the outer radius of the disc. In order to place a reasonable upper limit on the inclination we adopt a primary mass, $M_1\,=\,0.6\,\rmn{M}_\odot$, and mass ratio, $q\,=\,0.01$. We take 80 percent of the volume radius of the Roche lobe of the primary \citep{1983ApJ...268..368E} as an estimate of the maximum disc radius, giving an inclination, $i\,\le11 ^{\circ}$.

The measured separation of the peaks in the line profiles also places an upper limit on the velocity of the donor, $K_2 \le 160$\,km\,s$^{-1}$. If we assume that the measured radial velocity amplitude reflects the motion of the primary, $K_1$, we obtain a mass ratio, $q \ge 0.09$. This is larger than typically found in AM CVn binaries with similar orbital periods \citep[][and references therein]{2005MNRAS.361..487R,2006MNRAS.365.1109R,2013MNRAS.432.2048K,2012MNRAS.421.2310A}, and we expect that $K_1$ is in fact much smaller than our estimate. The radial velocity amplitude we obtained likely overestimates $K_1$ due to contamination by the higher velocity S-wave.

\citet{2006ApJ...640..466B} predict white dwarf surface temperatures in the range $\sim$13\,000 -- 27\,000\,K for an AM CVn binary with an orbital period of 35\,minutes; the similar period systems V406 Hya and SDSS\,J1240$-$0159 both have estimated white dwarf temperatures of 17\,000\,K \citep{2006ApJ...640..466B,2004RMxAC..20..254R}. Our estimate for SDSS\,J1730 is at the lower end of the expected range, however, at the orbital period of this system the disc may contribute significantly to the flux in quiescence \citep{2006ApJ...640..466B}, especially with the disc being so close to `face-on'. The temperature of the accreting white dwarf may therefore be higher than this estimate from the continuum. The spectra of SDSS\,J1730 show no sign of helium absorption from the accretor, a feature that has been observed in several AM CVn binaries with orbital periods of 30--40\,minutes \citep[e.g.][]{2006MNRAS.365.1109R}. This would not be unexpected for a white dwarf with a temperature of 12\,700\,K, for which the disc would easily hide the weak absorption lines; however, a hotter accretor would require that the disc is still relatively bright in comparison. 

Metals are often detected in the quiescent spectra of AM CVn binaries, and have been used to constrain the evolutionary history of some systems \citep{1991ApJ...366..535M,2010MNRAS.401.1347N}. In particular the presence of nitrogen lines without the detection of carbon/oxygen would suggest a helium white dwarf donor \citep{2010MNRAS.401.1347N}. With N\,\textsc{ii} and Ne\,\textsc{i} lines detected in the spectrum, we would expect to see strong carbon and oxygen features \citep[][found that large N/C and N/O ratios were required to suppress these lines]{1991ApJ...366..535M}. Our spectral range does not extend far enough into the red to cover the wavelengths of many of the strongest carbon, nitrogen and oxygen lines, so we cannot strongly constrain the N/C or N/O ratio with our data. That we see nitrogen lines but the C\,\textsc{ii} lines at 4267 and 6580\,\AA\ are missing, implies an overabundance of nitrogen that likely rules out all but the least-evolved helium star donors \citep{2008AstL...34..620Y,2010MNRAS.401.1347N}.

Models of helium accretion discs with solar metal abundances predict that Si\,\textsc{ii} 6347 and 6371, Fe\,\textsc{ii} 5169, and the Ca\,\textsc{ii} H and K lines should be the strongest metal lines \citep{1991ApJ...366..535M,2005MNRAS.361..487R}. Whilst the Si and Fe lines are clearly detected, there is no sign of Ca emission. This unexplained absence of calcium has been noted in several longer period AM CVn systems \citep[e.g.][]{2013MNRAS.432.2048K}, including the similar period system SDSS\,J1240$-$0159 \citep{2005MNRAS.361..487R}. However, Ca\,\textsc{ii} emission has been detected in the slightly shorter period systems PTF1\,J0943+1029, PTF1\,J0435+0029 \citep{2013MNRAS.430..996L} and V406 Hya \citep{2006MNRAS.365.1109R}, the latter of which has a spectrum that otherwise appears remarkably similar to SDSS\,J1240. \citet{2013MNRAS.432.2048K} discuss this anomaly in more detail, suggesting that sedimentation or settling could explain the under abundance of calcium. With the exception of the peculiar AM CVn system SDSS\,J0804+1616, there seems to be a sharp cut-off period for the appearance of calcium in AM CVn accretion discs.

The low inclination of SDSS\,J1730 causes the emission lines to be very narrow, making their identification much easier. In most systems the broad double-peaked lines blend together, making it difficult to determine which lines are present, and their relative strengths. This makes SDSS\,J1730 an excellent target for a detailed abundance analysis, which has proved difficult in other AM CVn binaries. We encourage $I$-band spectroscopic observations in order to target the carbon, nitrogen and oxygen lines expected in this region that can reveal the prior evolution of the binary \citep{2010MNRAS.401.1347N}.


\section{Conclusion}

We have presented time-resolved spectroscopy of the AM CVn binary SDSS\,J173047.59+554518.5, measuring an orbital period of 35.2\,$\pm$\,0.2\,minutes, confirming the ultra-compact binary nature of the system.

The average spectrum shows strong double-peaked helium emission lines, as well as a variety of metal lines, including neon, which has only been identified previously in GP Com. We detect no calcium in the accretion disc, this unexplained under-abundance of calcium has been noted in the majority of the longer-period AM CVn binaries.

We estimate the temperature of the continuum to be 12\,700\,$\pm$\,100\,K using a blackbody fit. The absence of obvious underlying absorption in the average spectrum is consistent with an accreting white dwarf at this temperature, however, the disc is expected to make a significant contribution to the continuum flux at the orbital period of SDSS\,J1730, which likely affects this estimate \citep{2006ApJ...640..466B}.

Using the measured peak velocities of the narrow double-peaked emission lines, we estimate an inclination, $i\,\le11 ^{\circ}$. This low inclination makes SDSS\,J1730 an excellent system for the identification of emission lines, and a good target for abundance analysis. With $I$-band spectroscopy we would be able to investigate the N/C and N/O ratios that reveal much about the prior evolution of AM CVn binaries \citep{2010MNRAS.401.1347N}.


\section*{Acknowledgements}

We thank the anonymous referee for useful comments and suggestions. PJC acknowledges the support of a Science and Technology Facilities Council (STFC) studentship. DS and TRM acknowledge support from the STFC grant no. ST/F002599/1. T. Kupfer acknowledges support by the Netherlands Research School for Astronomy (NOVA). GN acknowledges an NWO-VIDI grant.

Based in part on observations made with the William Herschel Telescope operated on the island of La Palma by the Isaac Newton Group in the Spanish Observatorio del Roque de los Muchachos of the Instituto de Astrofísica de Canarias.

Based in part on observations obtained under programme GN-2012A-Q-54 at the Gemini Observatory, which is operated by the Association of Universities for
Research in Astronomy, Inc., under a cooperative agreement with the
NSF on behalf of the Gemini partnership: the National Science
Foundation (United States), the Science and Technology Facilities
Council (United Kingdom), the National Research Council (Canada),
CONICYT (Chile), the Australian Research Council (Australia),
Minist\'{e}rio da Ci\^{e}ncia, Tecnologia e Inova\c{c}\~{a}o (Brazil)
and Ministerio de Ciencia, Tecnolog\'{i}a e Innovaci\'{o}n Productiva
(Argentina).

Funding for SDSS-III has been provided by the Alfred P. Sloan Foundation, the Participating Institutions, the National Science Foundation, and the U.S. Department of Energy Office of Science. The SDSS-III web site is http://www.sdss3.org/.

This research has made use of NASA's Astrophysics Data System and the NIST Atomic Spectra Database.


\label{lastpage}

\end{document}